\documentclass[twocolumn,english,aps,superscriptaddress,prl,floatfix,nofootinbib]{revtex4-2}
\usepackage[latin9]{inputenc}
\usepackage{color}
\usepackage{babel}
\usepackage{amsmath}
\usepackage{amssymb}
\usepackage{mathtools}
\usepackage{graphicx}

\newcommand{\jp}{J^{\prime}}

\graphicspath{{./}{./figs/}}

\begin{document}

\title{Unveiling Magnetic Frustration via the Elastocaloric Effect}

\author{Eric C. Andrade}
\affiliation{Instituto de F\'isica, Universidade de S\~ao Paulo, 05315-970, S\~ao Paulo, SP, Brazil}
\author{Pedro M. C\^onsoli}
\affiliation{Department of Physics, Arizona State University, Tempe, AZ 85287, USA}
\author{Matthias Vojta}
\affiliation{Institut f\"ur Theoretische Physik and W\"urzburg-Dresden Cluster of Excellence ctd.qmat, Technische Universit\"at Dresden, 01062 Dresden, Germany}
\begin{abstract}
Motivated by experimental progress in pressure and strain tuning of quantum materials, we examine the thermodynamic response of frustrated magnets to uniaxial strain. Specifically, we study Ising and Heisenberg models on spatially anisotropic triangular (and, for the Ising model, also kagome) lattices. We determine the entropy as a function of temperature and strain, and use it to compute the elastic Gr\"uneisen ratio $\eta$. The Ising models can be strain-tuned into and out of classical spin-liquid phases, and we show that $\eta$ can become arbitrarily large at low temperature $T$ near the point of maximal frustration, a universal hallmark of an extensive ground-state entropy. In contrast, the spin-$1/2$ Heisenberg model is moderately frustrated and displays multiple $T=0$ phase transitions. These transitions dominate $\eta$ at low $T$ while the intermediate-$T$ behavior is similar to that of the Ising model. We discuss the extent to which the elastic Gr\"uneisen ratio can be used to deduce the phase diagram, and we connect our results to recent experiments on triangular-lattice magnets.
\end{abstract}

\date{\today}
\maketitle


Controlling the properties of quantum materials by pressure and strain has become a major field of condensed matter research, in pursuit of both novel phenomena and functionalities. Such tuning by pressure and strain is appealing because it is typically reversible and does not introduce quenched disorder, unlike chemical doping. Moreover, uniaxial pressure enables controlled changes in symmetry and can thus serve as a nontrivial probe of many-body physics \cite{hicks_rev}. In this context, recent examples of uniaxial-pressure studies include the control of crystal-field levels in magnetic compounds \cite{gati23,zic24}, the enhancement of the lattice Kondo effect \cite{panja24}, the modification of multi-component superconducting states \cite{hicks17,ikeda21,klauss21,li22,ghosh24}, the exploration of critical properties of altermagnets \cite{chakraborty24,takahashi25,ohlendorf26}, and the tuning of frustration in a triangular-lattice organic magnet \cite{lieberich25}.

In fact, highly frustrated magnets \cite{lacroix11,castelnovo12,starykh15,vojta18} constitute a particularly interesting arena for uniaxial-pressure tuning, as large geometric frustration often relies on a delicate balance of exchange interactions protected by lattice symmetries. As a result, such systems are highly sensitive to symmetry-changing deformations, and novel phases can emerge \cite{chen86,alicea09,weichselbaum11,powell11,yunoki06,heidarian09,ghorbani16,gonzalez20,gonzalez22}.

An experimental challenge is to properly probe materials under uniaxial strain, as conventional thermodynamic measurements and many spectroscopic techniques cannot be readily applied in a pressure-cell environment. Among the various experimental advances over the last decade are accurate low-temperature measurements of the elastocaloric effect, where the pressure-induced temperature change is monitored under adiabatic conditions. The quantity of interest is the elastic Gr\"uneisen ratio $\eta$, the ratio of the entropy change under strain and the specific heat \cite{ikeda19}. This Gr\"uneisen ratio is known as a viable tool to detect quantum phase transitions, as it displays a universal divergence at quantum critical points \cite{garst03}.
The recent experiment of Ref.~\onlinecite{lieberich25} has demonstrated control over geometric frustration and its detection via elastocaloric measurements. Still, a comprehensive understanding of the features seen or expected in such measurements is lacking.

The purpose of the present letter is to fill this gap:
As a prime example, we investigate highly frustrated anisotropic Ising models under uniaxial deformations, revealing a rich interplay between classical spin-liquid behavior and symmetry-breaking order \cite{chen86}. We compute their thermodynamic properties using closed-form analytic solutions. At the isotropic point, the system displays an extensive entropy at temperature $T=0$,  which is quenched by any strain $\epsilon$.  As a result, the elastocaloric response $\eta$ -- which is a smooth function of $T$ and $\epsilon$ for $T>0$ that vanishes at the isotropic point -- becomes giant: At small strain $|\epsilon|$, $\eta$ is exponentially large at low $T$, signaling the proximity of the classical spin liquid.
We compare these findings with those for the less-frustrated triangular-lattice spin-$1/2$ Heisenberg model. In the latter, the intermediate-$T$ behavior is qualitatively similar to that of the Ising model and can therefore be broadly interpreted as a measure of frustration. However, this correspondence breaks down at low $T$, where in the Heisenberg model phase transitions and associated quantum criticality -- which occur away from the isotropic point -- dominate. Ultimately, our computations serve as a theoretical framework to guide future experiments.

\begin{figure}[!bt]
\includegraphics[width=1\columnwidth]{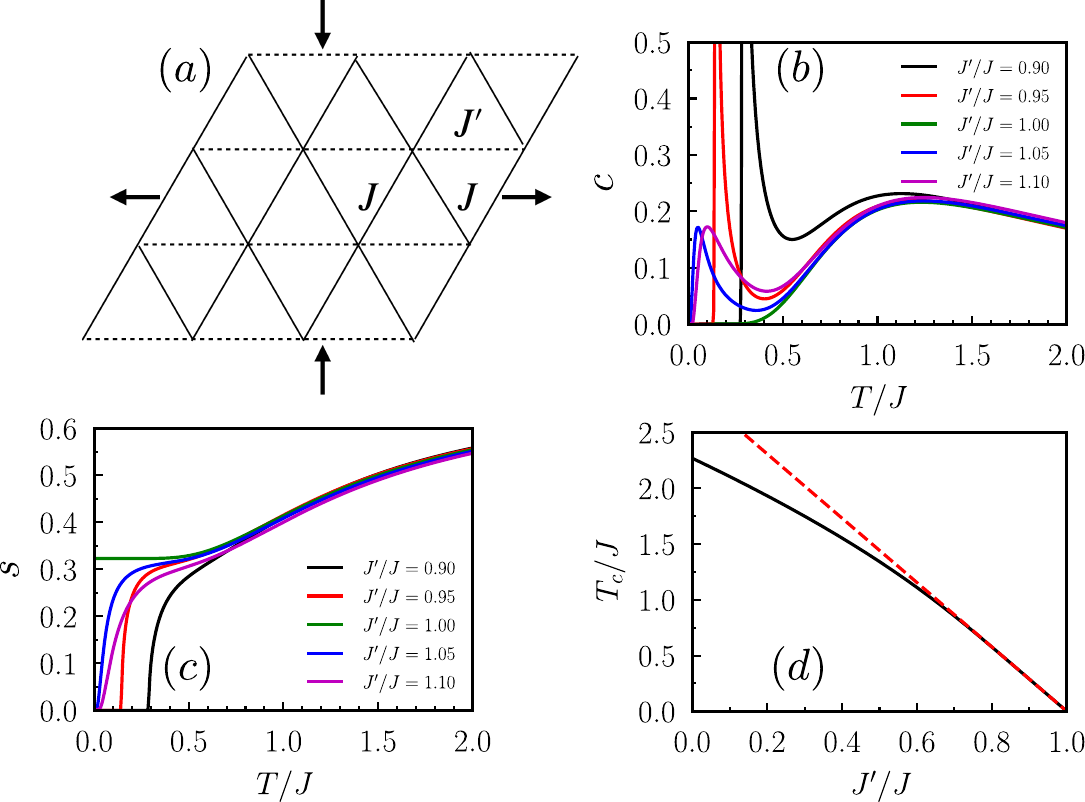}
\caption{\label{fig:tri_aniso}
(a) Spatially anisotropic triangular lattice, with exchange coupling $J$ along two directions (full lines) and $\jp$ along the remaining one (dashed lines). Arrows indicate the deformation induced by uniaxial pressure corresponding to positive strain $\epsilon$.
(b-d) Exact thermodynamic results for the Ising model,  with
(b) specific heat per site, $c$, as function of temperature $T$ for different values of $\jp$,
(c) entropy per site, $s$, as function of $T$ for several values of $\jp$, and
(d) critical temperature for $\jp/J < 1$ as function of $\jp$.
}
\end{figure}

\paragraph{Ising models.---}
We consider simple nearest-neighbor Ising models, with a spatial coupling anisotropy which reflects the influence of uniaxial pressure. We start with the case of the triangular lattice, with the Hamiltonian
\begin{align}
\mathcal{H} & =
 J\sum_{\left\langle ij\right\rangle }\sigma_{i}\sigma_{j}+J^{\prime}\sum_{\left\langle ij\right\rangle ^{\prime}}\sigma_{i}\sigma_{j},
\label{eq:h}
\end{align}
where we consider anisotropic antiferromagnetic couplings between nearest neighbors. The coupling $J$ acts along two of the three directions of the triangular lattice and $\jp$ along the remaining one, Fig.~\ref{fig:tri_aniso}(a). The Ising variables assume values $\sigma_i = \pm 1$ at each lattice site.

To study the general case,  we employ the exact solution of the model obtained by Houtappel \cite{houtappel50,baxter07} to write down the free energy per site:
\begin{align}
f & =-T\log2-T\frac{1}{8\pi^{2}}\int_{0}^{2\pi}d\omega_{1}\int_{0}^{2\pi}d\omega_{2}\log\left[C^{\prime}C^{2}-S^{\prime}S^{2}\right.\nonumber \\
 & +\left.S^{\prime}\cos\omega_{1}+S\left(\cos\omega_{2}+\cos\left(\omega_{1}+\omega_{2}\right)\right)\right]
\label{eq:houtappel}
\end{align}
where $C=\cosh 2\beta J$,  $S=\sinh 2 \beta J$,  $C^{\prime}=\cosh 2 \beta \jp$, and  $S^{\prime}=\sinh 2\beta \jp$.

This model interpolates between several well-known limits. In the maximally frustrated case $J = J'$, the system realizes a classical spin liquid. The ground state was determined by Wannier~\cite{wannier50}, who showed that the system exhibits an extensive degeneracy, with a residual-entropy density of $S_0/N = s_0 = 0.323066$, and no thermal phase transition. In the limit $J' = 0$, the lattice reduces to the square-lattice Ising model, which undergoes a finite-temperature phase transition at $T_c/J \simeq 2.27$ \cite{onsager44}, with the ground state corresponding to an antiferromagnetic N\'eel order. Finally, when $J = 0$, the system decouples into independent Ising chains and shows no long-range order at finite $T$. Sample results are shown in Fig.~\ref{fig:tri_aniso}.

For $\jp > J$, the system consists of antiferromagnetic chains along the direction of $\jp$ with frustrated inter-chain coupling; this results in a subextensive residual entropy proportional to the linear system size $L$ \cite{stephenson70}. The thermal crossover to a state with a large entropy of order $N s_0$ occurs at a low temperature $T_0 \approx \left(\jp -J\right) $. There is no long-range order for finite $T$ in this case, and the specific heat shows a peak at $T_0 $,  Fig.~\ref{fig:tri_aniso}(b).

Conversely, antiferromagnetic order is realized for $\jp < J$, as the frustrated bond of each triangle now lies along $\jp$. The specific heat in Fig.~\ref{fig:tri_aniso}(b) displays a pronounced peak marking the transition, with the ordering temperature given by $T_c = 2\left(J - \jp\right)/\ln 2$ close to the maximally frustrated case. As $\jp \to 0$, we recover the ordering temperature of the square lattice, Fig.~\ref{fig:tri_aniso}(d).

The entropy as a function of $\jp/J$ and temperature is shown in Fig.~\ref{fig:tri_aniso}(c). At the isotropic point, we have an extensive low-$T$ entropy which is immediately quenched at a temperature scale of order $ \left| J - \jp \right|$, regardless of the development of long-range order. This indicates a strong sensitivity of the system towards strain, as small changes will produce a large thermodynamic response at low $T$.


\paragraph{Elastocaloric effect.---}
The elastocaloric effect measures the adiabatic temperature change, $\Delta T$, upon application of strain, $\Delta\epsilon$. The resulting Gr\"uneisen ratio $\eta$ can be linked to strain-induced entropy variations by
\begin{align}
\eta & =
 \left(\frac{\Delta T}{\Delta\epsilon}\right)_{s}=-\frac{T}{c_\epsilon}\left(\frac{\partial s}{\partial\epsilon}\right)_{T},
\label{eq:eta}
\end{align}
where $s$ and $c_\epsilon$ are the entropy and specific heat per site, respectively. In particular, $\eta = 0$ at points where the entropy is extremal.

Empirically,  electron hopping amplitudes change under strain because the wave-function overlaps are altered. This leads to an exponential dependence of the exchange interaction on bond lengths, as is characteristic of superexchange mechanisms. In the small-strain limit, the dependence of the exchange interaction on the strain parameter $\epsilon$ can be linearized as follows \cite{ribeiro09,  rachel16}:
\begin{equation}
	J = J_{0} + \alpha \epsilon, ~~
	J^{\prime} = J_{0} - \epsilon,
	\label{eq:exc_strain}
\end{equation}
where $J_{0}$ is the exchange coupling in the uniform (unstrained) limit, employed below as the unit of energy and temperature. $\alpha$ is a parameter that encodes the bond geometries and the magnetoelastic coupling. Experimentally, uniaxial pressure $p$ will lead to strain both along and perpendicular to the direction of the pressure, according to the Poisson ratio. Here we implicitly assume $\epsilon\propto p$ and treat $\alpha>0$ as a free parameter.

\begin{figure}[!tb]
\includegraphics[width=1\columnwidth]{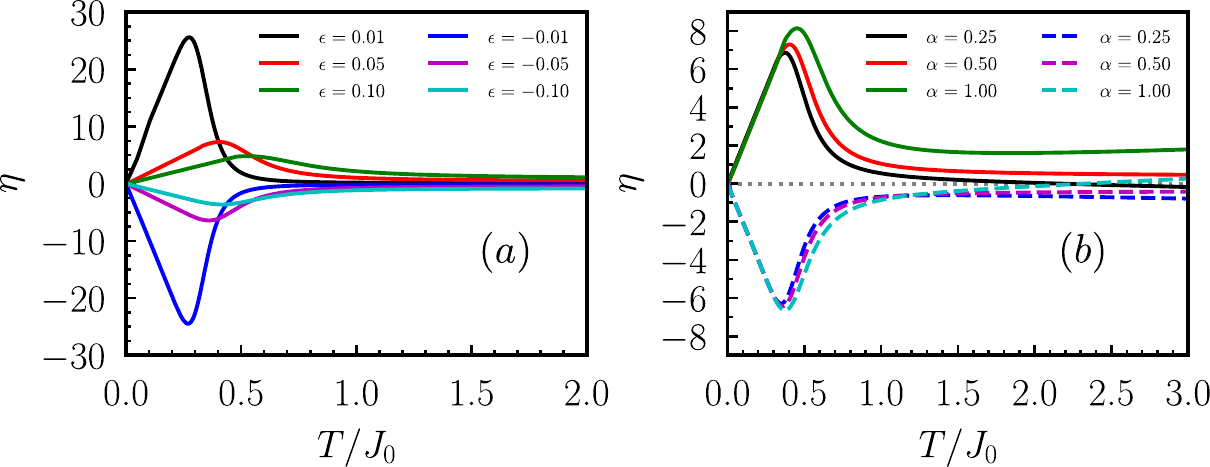}
\caption{\label{fig:tri_eta}
Elastocaloric effect $\eta$ for the anisotropic triangular-lattice Ising model as a function of temperature $T$.  We employ the parametrization for the exchange constant as in Eq.~\eqref{eq:exc_strain}.  (a) $\eta$  as a function of $T$ for $\alpha=0.5$ and several values of the strain $\epsilon$.  (b) $\eta$ as a function of $T$ for $|\epsilon|=0.05$ and several values of $\alpha$.  The full (dashed) curves correspond to $\epsilon > (<) 0$.
}
\end{figure}

Fig.~\ref{fig:tri_eta} presents the results for the triangular-lattice Ising model. At the isotropic point, the entropy reaches a maximum, which consequently leads to a sign change of the elastocaloric effect. The sign of $\eta$ is determined by the fact that the low-$T$ entropy decreases under applied strain, such that $\eta$ is positive for $\epsilon > 0$ (tensile strain) and negative for $\epsilon < 0$ (compressive strain).  For higher temperatures,  $\eta$ can change sign depending on the values of $\epsilon$ and $\alpha$.


\begin{figure}[!b]
\includegraphics[width=1\columnwidth]{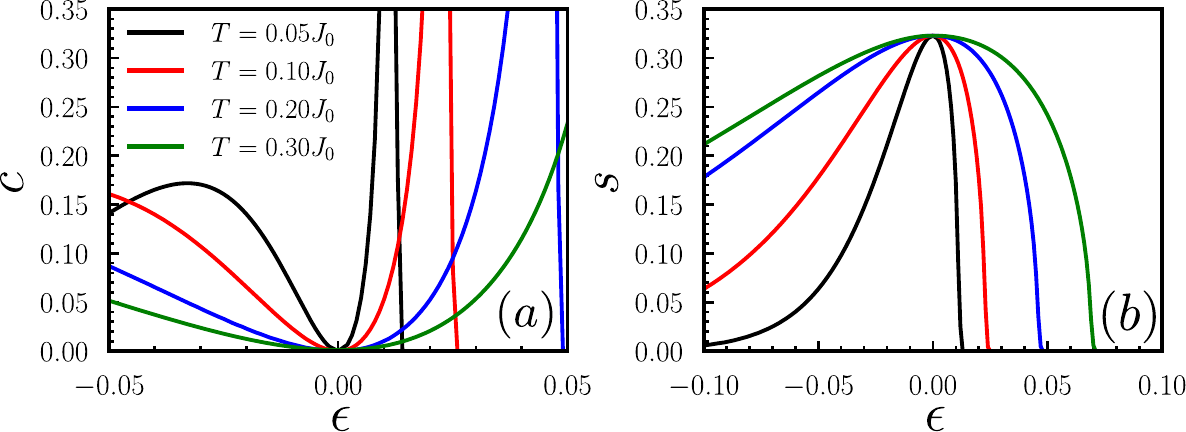}
\caption{\label{fig:tri_cs}
Strain dependence of (a) specific heat $c$ and (b) entropy per site $s$ at low temperatures $T$ for the anisotropic triangular-lattice Ising model, computed with $\alpha=0.5$. The singular behavior at finite positive $\epsilon$ arises from the thermal phase transition. The color coding is the same for both figures.
}
\end{figure}

\begin{figure}[!tb]
\includegraphics[width=0.9\columnwidth]{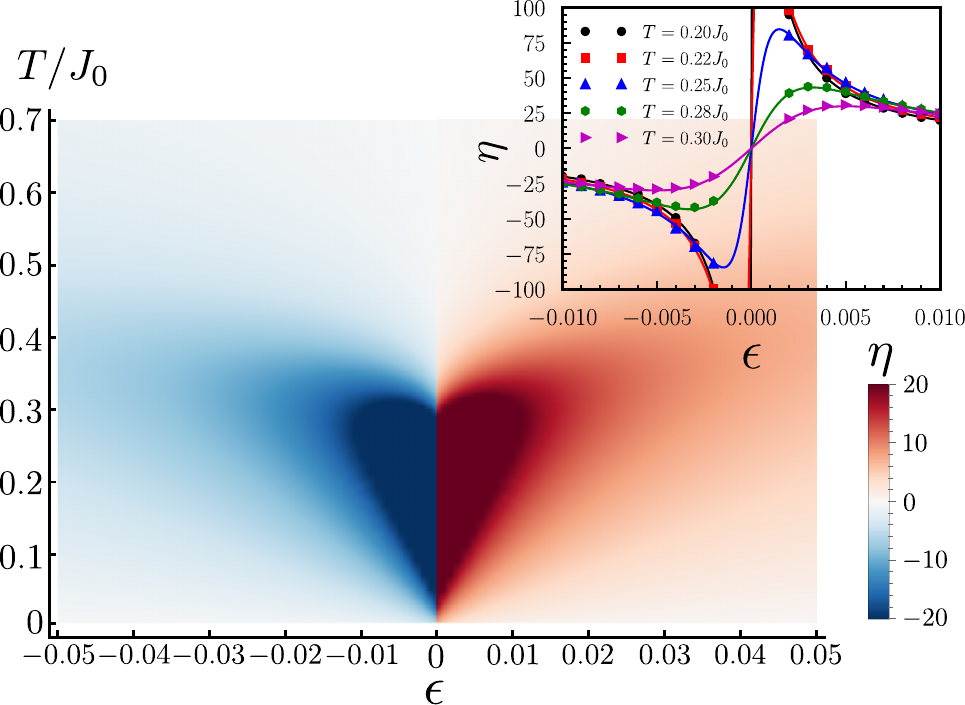}
\caption{\label{fig:tri_color}
Elastocaloric effect $\eta$ for the anisotropic triangular-lattice Ising model as a function of temperature $T$ and uniaxial strain $\epsilon$, computed with $\alpha=0.5$. The color code indicates the value of $\eta$.  We impose cutoffs $\eta= \pm20J_0$ for the sake of presentation. The inset shows cuts $\eta(\epsilon)$ at low $T$, illustrating the non-commuting limits. The points represent the exact computation of $\eta$, while the lines depict the low-$\epsilon$ expansion around the isotropic point, as discussed in the text.
}
\end{figure}

The elastocaloric response is significantly enhanced as $|\epsilon| \to 0$ in the low-$T$ regime, $T\lesssim 0.3 J_0$, where finite strain induces a large drop in $s$. Fig.~\ref{fig:tri_cs} illustrates the strain dependence of both specific heat and entropy as function of strain $\epsilon$ at low $T$: Both are analytic functions at small $\epsilon$ which can be expanded as $c=c_0+c_2 \epsilon^2$ and $s=\tilde{s}_0-s_2\epsilon^2$, with $c_0\propto\exp(-\Delta/T)$ the specific heat at the isotropic point with $\Delta\propto J_0$ a gap parameter, $\tilde{s}_0$ being the residual-entropy density $s_0$ plus an exponentially small finite-$T$ correction, and $c_2$ and $s_2$ are linked as $2s_2=c_2$ and scale as $1/T^2$.
Upon taking the $T\to0$ limit at finite $\epsilon$ we find $\eta = T/\epsilon$, as nicely seen in Fig.~\ref{fig:tri_eta}. Remarkably, this behavior holds regardless of the presence of a thermal phase transition. In contrast, if we take the limit $\epsilon\to0$ at finite $T$, we obtain $\eta = \epsilon T (2s_2)/ c_0$. Hence, $\eta\propto\epsilon$ with a slope that is exponentially large at small $T$. The two limits do not commute, and the crossover happens along a line given by $\epsilon^2=c_0/c_2$, hence exponentially close to the isotropic point. Therefore, $\eta$ becomes exponentially large at low $T$, as illustrated in the inset of Fig.~\ref{fig:tri_color}. We identify this as a hallmark of classical models with extensive ground-state entropy, as it also holds for the anisotropic Ising model on the kagome lattice; see End Matter.
In contrast, we will demonstrate below for the Heisenberg model that its low-$T$ response is continuous and reflects the model's specific details.



The low-$T$ profile of $\eta(T)$ is approximately antisymmetric with respect to the strain $\epsilon$, see Figs.~\ref{fig:tri_eta}(a) and \ref{fig:tri_color}. This indicates that the elastocaloric response is primarily governed by the distance from the isotropic point, despite the fundamentally distinct physics in the two strain regimes: the $\epsilon > 0$ region exhibits long-range order at low temperatures, whereas the $\epsilon < 0$ region does not.
We also find that the maxima of $\left| \eta (T) \right|$ are not associated with thermal phase transitions, as they occur in the temperature window between the two specific-heat peaks, see Fig.~\ref{fig:tri_aniso}(b) and Fig.~\ref{fig:tri_color}. The higher-temperature peak in $c(T)$, located at $T \sim J$, reflects the enforcement of the local constraint that two spins on each triangle must point opposite to the third. By contrast, the lower-temperature peak at $T \sim \left|J - \jp \right|$ marks the temperature below which the residual entropy of the isotropic point is quenched.
%

This discussion highlights that $\eta$ is insensitive to thermal phase transitions. This observation is consistent with earlier studies on the Gr\"uneisen ratio, which demonstrated that while the ratio diverges at a quantum critical point, it remains finite at a thermal phase transition \cite{garst03}. (We note, however, that the Gr\"uneisen ratio may be significantly enhanced near a thermal critical endpoint \cite{bartosch10}.)
According to Eq.~\eqref{eq:eta}, for the elastocaloric effect to remain finite at the critical temperature $T_c$, both the strain derivative of the entropy and the specific heat must exhibit the same divergence. This requirement is fulfilled in our model: near $T_c$, both quantities diverge as the second derivative of the free energy. In this critical region, we obtain $\eta = (T/T_c)(\partial T_c/\partial \epsilon)$. Because $T_c$ depends smoothly on the strain $\epsilon$, as shown in Fig.~\ref{fig:tri_aniso}(d),  we obtain a regular, non-divergent elastocaloric response.

We applied the same type of analysis to investigate the elastocaloric effect in the antiferromagnetic Ising model on an anisotropic kagome lattice. As shown in End Matter, the results are remarkably (and perhaps surprisingly) similar to those of the triangular-lattice system. This indicates that the main thermodynamic signatures discussed above are universal to Ising models with points of extensive ground-state degeneracy.



\paragraph{Triangular-lattice Heisenberg model.---}
So far, we have illustrated how the elastic Gr\"uneisen ratio $\eta$ is affected by the accumulation of extensive entropy at highly-frustrated points in Ising models. Given that magnetic interactions in real materials usually exhibit sizable spin-flip terms beyond the Ising limit, we extend our analysis to a spin-$1/2$ Heisenberg model
\begin{align}
\mathcal{H} & =
 J\sum_{\left\langle ij\right\rangle }\vec S_{i} \cdot \vec S_{j}+J^{\prime} \sum_{\left\langle ij\right\rangle ^{\prime}} \vec S_{i}\cdot\vec S_{j}
\label{eq:h3}
\end{align}
defined on the same anisotropic triangular lattice shown in Fig.~\ref{fig:tri_aniso}(a). At $T=0$, this model realizes at least three different phases \cite{yunoki06,heidarian09,ghorbani16,gonzalez20,gonzalez22}: collinear N\'eel order for small $J'/J<1$, noncollinear spiral order around the isotropic point $J'=J$, and a quasi-one-dimensional spin liquid with algebraic correlations along $J'$ chains for large $J'/J>1$. The locations of the corresponding quantum phase transitions have been estimated as $J'/J\approx 0.6\ldots0.7$ \cite{ghorbani16,gonzalez22} and $J'/J\approx 1.4\ldots1.7$ \cite{heidarian09,ghorbani16,gonzalez22}, respectively. It has also been suggested that $\mathbb{Z}_2$ quantum spin liquids may exist on both the small-$J'/J$ \cite{yunoki06,heidarian09,ghorbani16} and large-$J'/J$ \cite{gonzalez20,gonzalez22} sides of the spiral phase.
Since the Hohenberg-Mermin-Wagner theorem forbids long-range magnetic order at $T>0$ \cite{hohenberg67,mermin66}, and the putative spin liquids are connected to the high-temperature paramagnetic phase by crossovers \cite{senthil00}, the model is devoid of thermal phase transitions. Thus, the only singularities occur at $T=0$ and, in contrast to the Ising models studied above, none of them coincide with the isotropic point.

\begin{figure}[!bt]
\includegraphics[width=1\columnwidth]{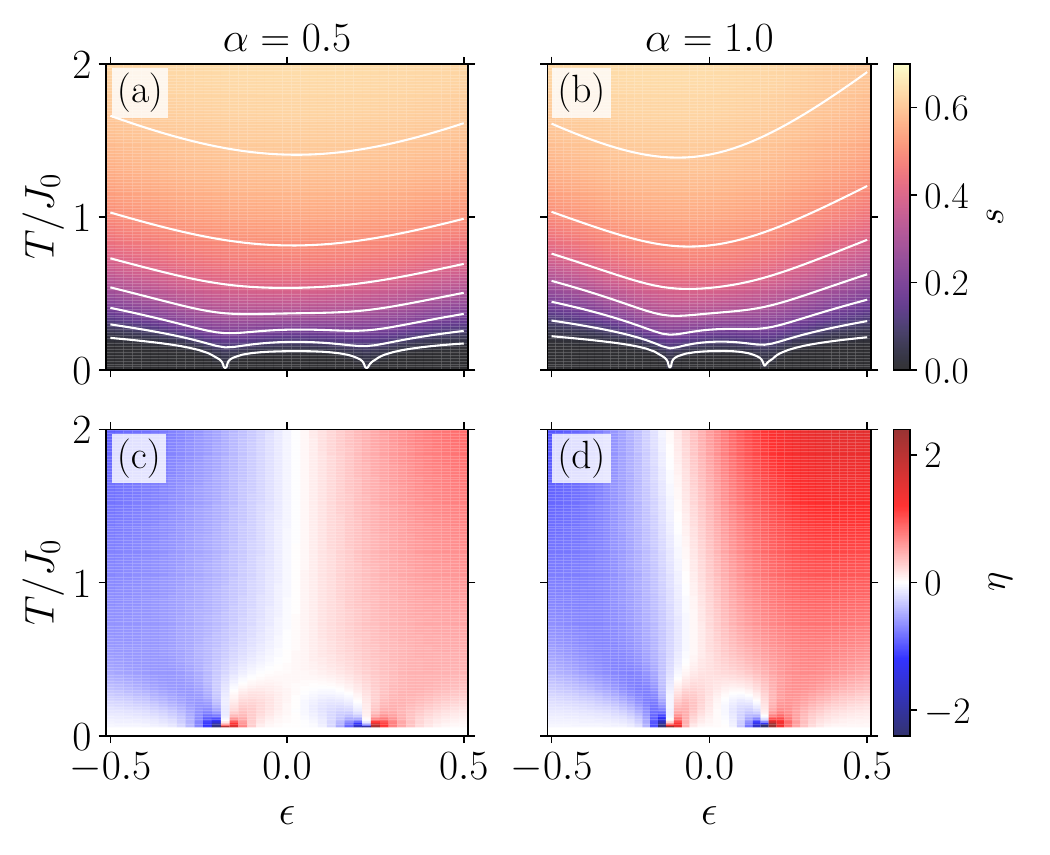}
\caption{\label{fig:heis_eta}
Thermodynamic observables computed by ED of the spin-$1/2$ Heisenberg model on the anisotropic triangular lattice. The left and right columns show data for anisotropy parameters $\alpha=0.5$ and $1.0$, respectively.
(a,b) Entropy per site, $s$, as a function of temperature $T$ and uniaxial strain $\epsilon$. The white curves correspond to $s = 0.03, 0.1, 0.2, \ldots, 0.6$. (c,d) Gr\"uneisen ratio $\eta$, showing signatures of two quantum phase transitions at low $T$ and a single characteristic sign change for $T/J_0 \gtrsim 0.5$. The data for $T/J_0 \le 0.05$ are omitted due to finite-size effects.
}
\end{figure}

We compute the thermodynamic properties of Eq.~\eqref{eq:h3} by performing exact diagonalization (ED) on a rhombic 18-site cluster subject to periodic boundary conditions. This cluster decomposes into three $J'$ chains of length six in the large-$J'/J$ limit and, crucially, it is commensurate with the $120^\circ$ order of the $J'=J$ ground state \cite{cluster_note}. To achieve an efficient numerical implementation of the problem, we used the global U(1) and translational symmetries of the Hamiltonian.

Results for the entropy and the elastic Gr\"uneisen ratio are shown in Fig.~\ref{fig:heis_eta}. At temperatures $T \gtrsim 0.5 J_0$, the behavior of $\eta$ resembles that of the Ising model, as there is a single $\eta=0$ line corresponding to the maxima of $S(\epsilon)$ at different $T$. Figures~\ref{fig:heis_eta}(c,d) demonstrate that the slope of this line is determined by the anisotropy parameter $\alpha$ and can take positive or negative values. In addition, we find that, when extrapolated to the limit $T\to 0$, the $\eta=0$ line terminates near $\epsilon=0$, as one might have anticipated from the fact that the classical analog of Eq.~\eqref{eq:h3} is maximally frustrated for $J'=J$.

At lower temperatures $T \lesssim 0.5 J_0$, we observe significant differences with respect to the Ising model. Because Eq.~\eqref{eq:h3} does not display (sub)extensive ground-state degeneracies, the interval spanned by $\eta$ is much smaller in Figs.~\ref{fig:heis_eta}(c,d) than in Fig.~\ref{fig:tri_color}. Furthermore, and most importantly, $\eta$ now shows multiple sign changes along fixed-$T$ cuts due to the presence of two quantum critical points at $J'/J \approx 1.34$ ($\epsilon<0$) and $J'/J \approx 0.70$ ($\epsilon>0$). We interpret these as transitions between the quasi-one-dimensional critical phase, noncollinear spiral order, and collinear N\'eel order as a function of increasing $\epsilon$.
Note that, while the analysis of Ref.~\onlinecite{garst03} implies that $\eta$ should diverge at such quantum critical points, our ED results do not show an actual divergence due to finite-size effects. Similarly, the absence of the aforementioned $\mathbb{Z}_2$ quantum spin liquids may be a finite-size artifact. Finally, the fact that the quantum critical points in Fig.~\ref{fig:heis_eta} are located almost symmetrically around the isotropic point induces a \textit{minimum} (instead of a maximum) of the entropy $s(\epsilon)$ near $\epsilon=0$ at low $T$, as visible in Fig.~\ref{fig:heis_eta}(a,b).


\paragraph{Summary.---}
We have evaluated the elastic Gr\"uneisen ratio for spatially anisotropic Ising models on both the triangular and the kagome lattice as well as for the triangular-lattice spin-$1/2$ Heisenberg model. We have examined the hypothesis that the strain dependence of the Gr\"uneisen ratio, $\eta(\epsilon)$, at fixed temperature can be used to locate the point of maximum frustration as characterized by an accumulation of entropy. This turns out to be generally true for the Ising models, for which -- in addition -- the presence of classical spin liquids induces a giant (exponentially large) response near the isotropic point at low $T$. In contrast, for the Heisenberg model, the sign change of $\eta(\epsilon)$ associated with an entropy maximum only occurs near the isotropic point at intermediate to high temperatures. At low $T$, the behavior of $\eta$ is instead dominated by quantum phase transitions of the spin state, all of which happen away from the isotropic point. Moreover, our results demonstrate that the temperature dependence of the Gr\"uneisen ratio can in general not be used to deduce thermal phase transitions.

Our study serves as a guide to ongoing and future experiments. One case in point is Ref.~\onlinecite{lieberich25}, where the organic magnet $\kappa$-(ET)$_2$Cu$_2$(CN)$_3$ was investigated under uniaxial pressure: Our results suggest that the association of finite-$T$ features of $\eta$ with continuous thermal phase transitions may not be straightforward. Other interesting frustrated magnets to consider are SrCu$_2$(BO$_3$)$_2$ \cite{jaime12,  zayed17,  shi22}  Ca$_{10}$Cr$_7$O$_{28}$ \cite{balz16,balz17,sonnenschein19}, and PbCuTe$_2$O$_6$ \cite{chillal20,hanna21}.
On the theory front, numerical studies of other types of transitions in frustrated magnets, e.g., into valence-bond solids and between spin liquids, would be desirable to obtain a more complete picture of the finite-$T$ behavior of the elastocaloric effect.


\acknowledgments

We thank F. Alarcon Acevedo, E. Gati, F. K\"ohler, and L. Janssen for discussions and collaborations on related topics.
We acknowledge support by the DFG through SFB 1143 (project id 247310070) and the W\"urzburg-Dresden Cluster of Excellence on Complexity, Topology and Dynamics in Quantum Matter---\textit{ctd.qmat} (EXC 2147, project id 390858490).
ECA was supported by CNPq (Brazil), Grant No. 302823/2022-0, and FAPESP (Brazil), Grant Nos. 2021/06629-4 and 2022/15453-0.  ECA also acknowledges the hospitality of TU Dresden, where part of this work was performed.
Part of PMC's work was supported by the U.S. Department of Energy, Office of Science, Office of Basic Energy Sciences, Material Sciences and Engineering Division under Award Number DE-SC0025247.


\appendix

\begin{figure}[!b]
\includegraphics[width=1\columnwidth]{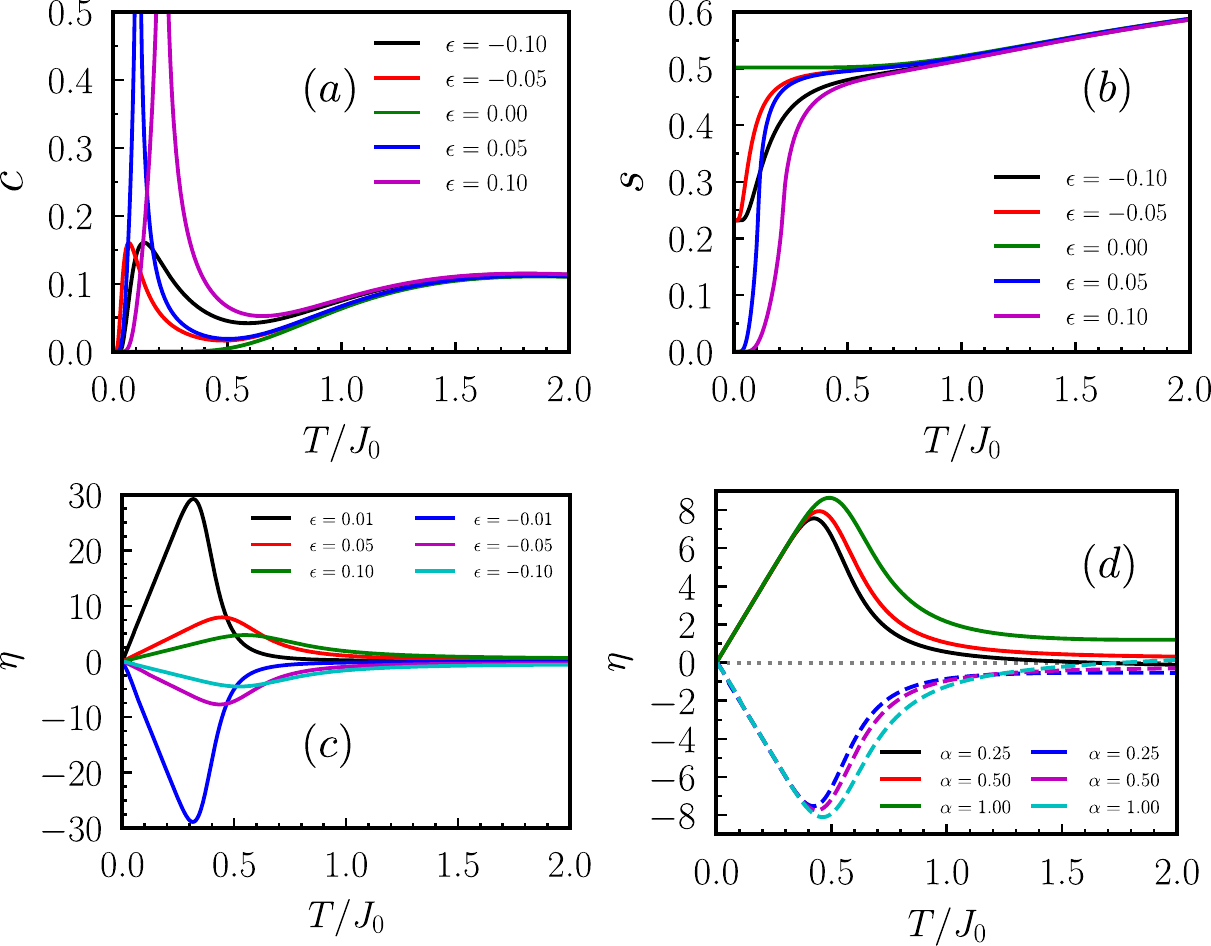}
\caption{\label{fig:kag_aniso}
Results for the Ising model on the spatially anisotropic kagome lattice, with the exchange $J$ running along two directions and the coupling $J'$ running along the remaining one,  as in Fig. ~\ref{fig:tri_aniso}(a). We implement the exchange parametrization given in Eq.~\eqref{eq:exc_strain}. In (a-c) we use $\alpha=0.5$.
(a) Specific heat per site, $c$, as function of temperature $T$ for different values of the strain $\epsilon$.
(b) Entropy per site, $s$, as function of $T$ for several values of $\epsilon$.
(c) Elastocaloric effect, $\eta$, as function of $T$ for several values of $\epsilon$.
(d) $\eta$ as a function of $T$ for $|\epsilon|=0.05$ and several values of $\alpha$. The full (dashed) curves correspond to $\epsilon > (<) 0$.
}
\end{figure}

\section*{END MATTER}

\paragraph{Kagome-lattice Ising model.---}
We illustrate the generality of the results for the classical model, also studying the anisotropic Ising model in Eq.~\eqref{eq:h} in the kagome lattice with a similar setup as in Fig. ~\ref{fig:tri_aniso}(a).  The free energy in this case reads \cite{kano53,diep04,li10}
\begin{align}
f & =\frac{-T}{24\pi^{2}}\int_{0}^{2\pi} \!\!\! d\omega_{1} \int_{0}^{2\pi} \!\!\! d\omega_{2}\log\left\{ 32S^{\prime}\left[S^{\prime}C^{2}-C^{\prime}S^{2}\right]\cos\omega_{1}\right.
\nonumber \\
 & +32S^{2}C\left[C^{\prime}-S^{\prime}\right]\left[\cos\omega_{2}-\cos\left(\omega_{1}-\omega_{2}\right)\right]\nonumber \\
 & +\left.16\left[\left(C^{\prime}C^{2}-S^{\prime}S^{2}\right)^{2}+C^{\prime2}+2C^{2}\right]\right\} ,\label{eq:kagome}
\end{align}
with the same notation as in Eq.~\eqref{eq:houtappel}.

In the isotropic limit, this model displays an extensive ground-state degeneracy with $s_0=0.502$.  For $\jp < J$, the system again displays long-range order, with the critical temperature $T_c \simeq \left(J-\jp\right)/\log2$ close to the isotropic point.  For $\jp > J$, on the other hand,  the system once more does not show long-range order,  retaining now a finite ground-state entropy of $s_0=0.231$.  The specific heat in Fig.~\ref{fig:kag_aniso}(a) and the entropy in Fig.~\ref{fig:kag_aniso}(b) illustrate these trends.

Sample results for the elastocaloric effects are in Figs.~\ref{fig:kag_aniso}(c) and (d).  The response is enhanced near the isotropic point, and it is approximately symmetric with respect to it.  The maximum of $\eta$ falls between the high-$T$ peak of the specific heat -- located at $T \sim J$ and reflecting the enforcement of the local constraint that two spins on each triangle must point opposite to the third -- and the lower-temperature peak -- occurring around $T \sim \left|J - \jp \right|$ and signaling a phase transition or the partial quench of the ground-state entropy.  The similarity of the results on the kagome and triangular lattices indicates that the dominant mechanism for enhancing the elastocaloric response is entropy release as one moves away from the isotropic point, and that all other features of the system become subleading.


\begin{thebibliography}{99}
\bibitem{cluster_note}
We also performed ED calculations for rhombic clusters with dimensions $4\times 4$ and $3\times 6$. The latter differs from the $6\times 3$ cluster discussed in the main text in that it has six $J'$ chains of length three. The results for these clusters agree semiquantitatively at elevated temperatures, $T/J_0 \gtrsim 0.5$, but show qualitative differences at low $T$. These are due to the incommensurability of the $4\times 4$ cluster with the three-sublattice $120^\circ$ order and the fact that the $J'$ chains in the $3\times 6$ cluster are too short to develop a quasi-one-dimensional spin liquid. Hence, given the size limitations of our ED, we believe that the results for the $6\times 3$ cluster, Fig.~\ref{fig:heis_eta}, are closest to the actual thermodynamic limit.
\end{thebibliography}
\end{document}